\documentclass{iaus}
\usepackage{graphicx}

\title[IAU 249.~~Orbital stability criteria]
{Orbital stability of planets in binary systems: \\
A new look at old results}

\author[J. Eberle et al.]   %% give here short author list %%
{J. Eberle, M. Cuntz and Z. E. Musielak}

\affiliation{
Department of Physics, University of Texas at Arlington, \\
Arlington, TX 76019-0059, USA \\
email: {\tt cuntz@uta.edu, zmusielak@uta.edu} \\
}

\pubyear{2008}
\volume{xxx}  %% insert here IAU Symposium No.
\pagerange{119--126}
\jname{Title of your IAU Symposium}
\editors{A.C. Editor, B.D. Editor \& C.E. Editor, eds.}
\begin{document}

\maketitle

\begin{abstract}

About half of all known stellar systems with Sun-like stars consist of two or
more stars, significantly affecting the orbital stability of any planet in these
systems.  This observational evidence has prompted a large array of theoretical
research, including the derivation of mathematically stringent criteria for the
orbital stability of planets in stellar binary systems, valid for the
``coplanar circular restricted three-body problem".  In the following, we use
these criteria to explore the validity of results from previous theoretical studies.
\end{abstract}

\firstsection % if your document starts with a section,

\section{Introduction}

Observational evidence for the existence of planets in stellar binary
(and higher order) systems has been given by
\cite[Patience et al. (2002)]{pat02}, 
\cite[Eggenberger et al. (2004, 2007)]{egg04,egg07}, and others.
\cite[Eggenberger et al.]{egg07}, presented data for more
than thirty systems, mostly wide binaries, as well as several triple star
systems, with separation distances as close as 20~AU (GJ~86).
These findings are consistent with previous theoretical
results which showed that planets can successfully form in binary
(and possibly multiple) stellar systems
(e.g., \cite[Kley 2001]{kle01}, \cite[Quintana et al. 2002]{qui02}),
known to occur in high frequency in the local Galactic neighborhood
(\cite[Duquennoy \& Mayor 1991]{duq91}, \cite[Lada 2006]{lad06},
\cite[Raghavan et al. 2006]{rag06}).
More recently, \cite[Bonavita \& Desidera (2007)]{bon07} performed a statistical
analysis for binaries and multiple systems concerning the frequency of
hosting planets, leading to the conclusion that there is no significant
statistical difference between binary systems and single stars.  The fact
that planets in binary systems are now considered to be relatively common
is also implied by the recent detection of debris disks in various
main-sequence stellar binary systems using the {\it Spitzer Space Telescope}
(\cite[Trilling et al. 2007]{tri07}).  The research team observed
69 main-sequence binary star systems in the spectral range of A3 to F8. 

In our previous work, we studied the stability of both S-type and P-type
orbits in stellar binary systems, and deduced orbital stability limits for
planets (\cite[Musielak et al. 2005]{mus05}).  P-type orbits lie 
well outside the binary system, where the planet essentially orbits the
center of mass of both stars, whereas S-type orbits lie near
one of the stars, with the second star acting as a perturbator.
The limits of stability were found to depend on the mass ratio between
the stellar components.  This topic has recently been revisited by
\cite[Cuntz et al. (2007)]{cun07} and
\cite[Eberle et al. (2008a, 2008b)]{ebe08a,ebe08b},
who used the concept of Jacobi's integral and Jacobi's constant
(\cite[Szebehely 1967]{sze67}, \cite[Roy 2005]{roy05}) to deduce stringent criteria
for the stability of planetary orbits in binary systems for the special case of the
``coplanar circular restricted three-body problem".  These criteria are used to contest
previous results on planetary orbital stability in binary systems available in the
literature.

%%% *** Fig.1
%%%%%%%%%%%%%%%%%%%%%%%%%%%%%%%%%%%%%%%%%%%%%%%%%%%%%%%%%%%%%%%%%
\begin{figure}
\begin{center}
\includegraphics[width=2.9in]{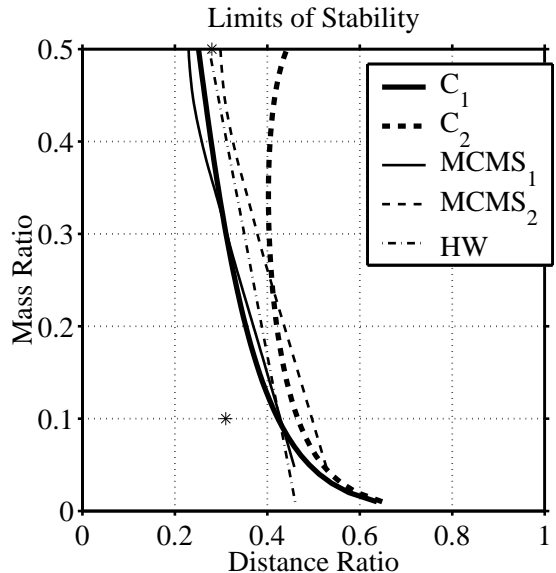}
\caption{
Limits of stability for planetary orbits for different mass
ratios $\mu$ between the secondary star and the sum of the two stellar
masses.  We show the result based on Jacobi's constant $C_1$
(thick solid line) (criterion of stability) and based on Jacobi's
constant $C_2$ (thick dashed line).  For comparison, we also
show the dividing lines between the regions of stability and marginal
stability (thin solid line) [MCMS$_1$] and between marginal
stability and instability [MCMS$_2$] (thin dashed line) previously
obtained by \cite{mus05}.  The two asterices ($\ast$) indicate the
stability limit by \cite{dav03} corresponding to $10^6$ orbits.
The stability limit from the earlier work by
\cite[Holman \& Wiegert (1999)]{hol99} [HW]
is depicted as a thin dash-dotted line.
}
\end{center}
\end{figure}
%%%%%%%%%%%%%%%%%%%%%%%%%%%%%%%%%%%%%%%%%%%%%%%%%%%%%%%%%%%%%%%%%

\section{Results and Conclusions}

By focusing on the coplanar circular restricted three-body problem, we
have been able to deduce absolute stability limits for small-mass planets,
which only depend on the mass ratio between the two stellar components
(see Fig.~1).  The analysis by \cite[Eberle et al. (2008a)]{ebe08a} elucidates
the physical backround of these limits.  They are closely related to the
so-called ``surfaces of zero velocity" (or zero velocity contours) that
form the boundaries of regions where the planet must be found.
Thus, planets are unable to escape from these regions, preventing them
from leaving the system entirely or from being captured by the other stellar
component.  This property is tantamount to orbital stability, although it does
not necessarily imply stability in the sense of quasi-periodicity.
The rationale of the contour is that it limits the allowable region
of the planet due to its limited available energy.

In the following, we compare the absolute stability limit of
\cite[Cuntz et al. (2007)]{cun07} and \cite[Eberle et al. (2008b)]{ebe08b}
to results from time-dependent orbital stability simulations
obtained by \cite[Holman \& Wiegert (1999)]{hol99},
\cite[David et al. (2003)]{dav03}, and
\cite[Musielak et al. (2005)]{mus05} (see Fig.~1).
The comparison of this newly found absolute stability limit with the
stability limits previously deduced by
\cite[Musielak et al.]{mus05} shows that their numerically derived
stability limit
largely agrees with the $C_1$-related stability criterion
(\cite[Cuntz et al. 2007]{cun07}, \cite[Eberle et al. 2008b]{ebe08b}),
although for a small range of mass ratios their stability criterion is too
strict.  In addition, there are discrepancies between our $C_2$ criterion for
unstable orbits and their criterion for marginally stable / unstable orbits.
Previous work by \cite[Holman \& Wiegert (1999)]{hol99} considered a large range
of eccentricities for the stellar binary components.  For circular orbits, their
stability limit is similar to the $C_1$-based stability limit in our study as well
as the stability limit of \cite[Musielak et al. (2005)]{mus05}, although a careful
analysis shows that small, but noticeable differences exist as the
\cite[Holman \& Wiegert]{hol99} criterion is somewhat less strict for mass ratios
between $\mu = 0.5$ and 0.1, but too strict for mass ratios of less than 0.1 ---
at least as viewed based on the polynomial fit given by the authors.

Other recent results have been given by \cite[David et al. (2003)]{dav03}.
They investigated
the range of orbital stability for a similar parameter range as previously
discussed.  Here the orbital stability of an Earth-mass planet around a
solar-mass star is studied in the presence of a companion star for a range
of companion masses between 0.001 and 0.5 $M_\odot$.
\cite[David et al. (2003)]{dav03} consider
both circular and elliptical orbits, and derive expressions for the expected
ejection time of the planet from the system.  For fixed companion masses,
the ejection time is found to be a steep function of the periastron distance
(or orbital radius for circular orbits) to the primary star.
\cite[David et al.]{dav03}
predict that the domain of orbital stability gets progressively smaller over
time according to the logarithm of the time of simulation, even though some
cases in their study do not seem to indicate this type of behavior.  Note
that their principle finding of a progressively decreasing stability limit
is fundamentally inconsistent with the result from the
\cite[Cuntz et al. (2007)]{cun07} and \cite[Eberle et al. (2008b)]{ebe08b}
studies that yield an analytically derived absolute limit of stability,
although those cases solely focus on the restricted three-body problem.
For $\mu = 0.3$ and zero eccentricity, \cite[David et al. (2003)]{dav03}
deduce a stability limit of $\rho_0 = 0.31 \pm 0.04$ for $10^6$ orbits,
which is well inside the analytically defined stability region obtained
in our study.

In conclusion, it is found that the comparison of the results
by \cite[Cuntz et al. (2007)]{cun07} and \cite[Eberle et al. (2008b)]{ebe08b} 
to those previously obtained by \cite[Holman \& Wiegert (1999)]{hol99},
\cite[David et al. (2003)]{dav03}, and \cite[Musielak et al. (2005)]{mus05},
show that their criteria for orbital stability are in good agreement
with our analytical criterion, although some small, but noticeable
differences exist.  For planets which do not fulfill our stringent
criterion of orbital stability, stability may still be possible
for a significant period of time.  However, in this case the outcome
must carefully be studied by employing long-term simulations and may
depend on various factors, such as, e.g., the mass and orbital starting
position of the planet.

\end{document}